# Confinement-driven inverse domain scaling in polycrystalline ErMnO$_3$

*Jan Schultheiß*[*,1], *Fei Xue*[2], *Erik Roede*[1], *Håkon W. Ånes*[1], *Frida H. Danmo*[1],
*Sverre M. Selbach*[1], *Long-Qing Chen*[2], *and Dennis Meier*[*,1]

[1] Department of Materials Science and Engineering, Norwegian University of Science and Technology, 7034, Trondheim, Norway
[2] Department of Materials Science and Engineering, The Pennsylvania State University, University Park, Pennsylvania 16802, USA

*corresponding authors: jan.schultheiss@ntnu.no; dennis.meier@ntnu.no

The research on topological phenomena in ferroelectric materials has revolutionized the way we understand polar order. Intriguing examples are polar skyrmions, vortex/anti-vortex structures and ferroelectric incommensurabilties, which promote emergent physical properties ranging from electric-field-controllable chirality to negative capacitance effects. Here, we study the impact of topologically protected vortices on the domain formation in improper ferroelectric ErMnO$_3$ polycrystals, demonstrating inverted domain scaling behavior compared to classical ferroelectrics. We observe that as the grain size increases, smaller domains are formed, which we relate to the interaction of the topological vortices with local strain fields. The inversion of the domain scaling behavior has far-reaching implications, providing fundamentally new opportunities for topology-based domain engineering and the tuning of the electromechanical and dielectric performance of ferroelectrics in general.

## 1. Introduction

Ferroelectric materials are the backbone of many electronic components, finding applications as capacitors, energy harvesters, and actuators.[1] The rich functionality of ferroelectrics is closely linked to their domain structure, and substantial contributions to their dielectric, piezoelectric and electromechanical responses originate from domain walls.[2,3] Because of the strong correlation between the domain morphology and the functional behavior, domain engineering is a powerful pathway for customizing the macroscopic performance of ferroelectrics.[4-6] Application of different electrode materials,[7] thickness variations,[8] and chemical doping[9] are common approaches that allow for stabilizing well-defined domain states on demand. Furthermore, strain effects are utilized to control the domains in ferroelectric single crystals and thin films, e.g., via substrate-related clamping effects[10] and microstructural engineering approaches that rely on dislocations,[11] secondary inclusions[12] or precipitates[13].

In polycrystals, an additional degree of freedom for microstructural engineering emerges from the three-dimensional (3D) confinement due to finite grain size.[14] A prominent example is the relation between the grain size, $g$, and the domain size, $d$, in BaTiO$_3$ polycrystals,[15-18] which facilitated the downscaling of multilayer ceramic capacitors.[19,20] The relation follows the universal power-law

$$d \sim g^m. \quad (1)$$

Importantly, for all ferroelectrics investigated so far, the exponent $m$ is positive, with a value of $m \approx 0.5$ for BaTiO$_3$[16] and Pb(Zr,Ti)O$_3$[21], indicating that smaller grains develop smaller domains. The driving force for this scaling behavior is elastic strain, which can be released via the formation of ferroelastic domain walls. With decreasing grain size the strained volume fraction increases,[22-24] requiring more domain walls for strain compensation so that smaller domains become energetically favorable.[25]

Here, we show that negative coefficients, $m$, arise in polycrystalline hexagonal manganites, demonstrating an inversion of the established grain-size-dependent scaling behavior of ferroelectric domains. Our systematic study of the relation between domain and grain sizes in ErMnO$_3$ shows that, in contrast with BaTiO$_3$ and Pb(Zr,Ti)O$_3$, the ferroelectric domains become smaller with increasing grain size, scaling with $m \approx -0.1$. Using phase field simulations, we relate the unusual behavior to topologically protected structural vortices that naturally form in the ferroelectric phase of ErMnO$_3$[26] and their interaction with elastic strain fields.[27-30] Our findings reveal the importance of topological defects for the electric long-range order in ferroelectric polycrystals, enabling anomalous domain scaling behaviors and conceptually new ways for domain engineering.

## 2. Geometrical 3D confinement in polycrystalline ErMnO$_3$

Ternary hexagonal manganites, $R$MnO$_3$ ($R$ = Sc, Y, In, and Dy – Lu), have been intensively studied as model system for improper ferroelectricity.[31,32] In contrast to proper ferroelectrics, such as LiNbO$_3$, Pb(Zr,Ti)O$_3$, and BaTiO$_3$, the spontaneous polarization, $P$, is not the primary order parameter.[33,34] In $R$MnO$_3$, the polarization arises as a symmetry enforced consequence of a structurally driven phase transition.[35-37] As a consequence, the material exhibits a large variety of unusual physical phenomena at the level of domains.[38,39] Intriguing examples range from domain walls with unique functional electronic properties[40-44] to topological vortex structures that have been utilized to test cosmological scaling laws[45-47]. Most of the research so far, however, focused on single crystals and thin films,[48,49] whereas only a few studies were performed on polycrystalline systems.[50-53] In particular, the impact of the microstructure on the ferroelectric domain distribution remains to be explored.

To systematically investigate the relation between the microstructure and domains, we synthesize high-quality polycrystalline ErMnO$_3$ from oxide precursors[54] with subsequent high-temperature heat-treatment (more details on processing can be found in the Supporting Information and microstructures are shown in Figure S1). To confirm the phase purity of our samples, we perform powder X-ray diffraction (XRD) of a crushed pellet (**Figure 1**a), confirming P6$_3$cm space group symmetry equivalent to ErMnO$_3$ single crystals.[55] Figure 1b presents the microstructure of the polycrystalline material with an average grain size of about 19.0 μm and the random orientation of the individual grains, measured by electron backscattered diffraction (EBSD, see Figure S2, Supporting Information for details). Figure 1c and 1d show piezoresponse force microscopy (PFM) images of the ferroelectric domain structure of the region presented in Figure 1b. The PFM images are recorded with a voltage amplitude of 10 V at a frequency of 40.13 kHz, displaying pronounced out-of-plane (Figure 1c) and in-plane (Figure 1d) contrast between the +$P$ and −$P$ domains (more details are provided in the Supporting Information). Within the different grains (grain boundaries are highlighted by white dashed lines) we observe the well-established domain structure of ErMnO$_3$,[40, 56, 57] consisting of ferroelectric 180° domains which come together in characteristic sixfold meeting points. These meeting points originate from the structural trimerization that drives the improper ferroelectric phase transition in hexagonal manganites and correspond to topologically protected vortex/anti-vortex pairs as explained elsewhere.[27, 56] Although the observed domain structures are qualitatively equivalent to those found in single crystals, Figure 1c and d also reveal an important difference: Isotropic domain distributions are only observed in certain regions, whereas the majority of the investigated area exhibits elongated stripe-like domains. The patterns are reminiscent of those reported to arise in single crystals under elastic strains.[29] It was demonstrated that elastic strain drives the vortex/anti-vortex pairs into opposite directions, unfolding the domain pattern into elongated stripe-like domains.[27, 28, 30] The PFM data thus reflects non-zero strain fields within the individual grains, which are indeed expected to arise due to the anisotropic thermal expansion of ErMnO$_3$ and grain-to-grain interactions.[58, 59] Interestingly, the PFM data also reveals that domain walls have a propensity to orient perpendicular to the grain boundaries (Figure 1e) as corroborated by the statistical evaluation presented in Figure S3 (Supporting Information). The domain walls, however, do not cross the grain boundaries (Figure 1f). This behavior is fundamentally different from the ferroelectric domain structures reported for polycrystalline BaTiO$_3$[60] and Pb(Zr,Ti)O$_3$[61], where domain walls tend to continue into neighboring grains.

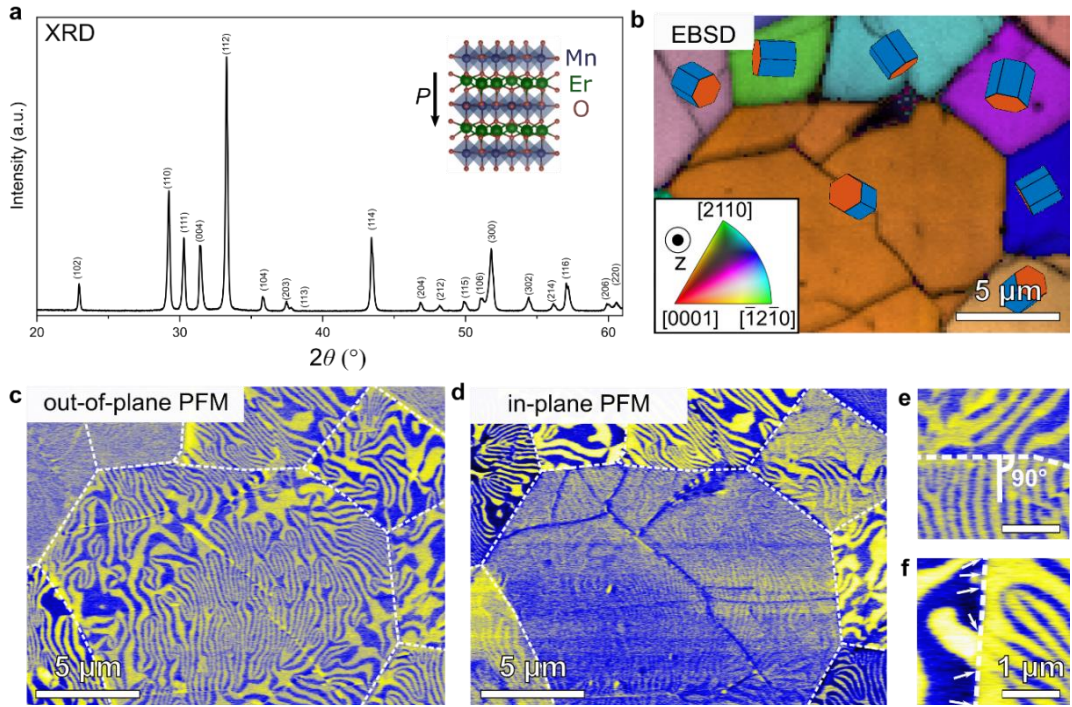

**Figure 1.** Structural and microstructural analysis of polycrystalline ErMnO$_3$. (a) XRD pattern of crushed polycrystalline ErMnO$_3$, showing that the material has a hexagonal structure with space group symmetry P6$_3$cm. The atomic structure of the ferroelectric phase of $R$MnO$_3$ with crystallographic data from ref. [55] is displayed in the inset. The direction of the spontaneous polarization vector, $P$, is indicated. (b) EBSD results showing the orientation of the grains of the polycrystalline material. The orientation of the individual unit cells is displayed schematically. Blue surfaces represent the $(10\bar{1}0)$ and red colors the $(0001)$ planes of the hexagonal crystal with space group P6$_3$cm. A detailed description of the fitting of the EBSD patterns is provided in Figure S3 (Supporting Information). (c,d) PFM out-of plane and in-plane images of the same area as presented in (b). Yellow and blue regions correspond to $\pm P$ domains. Dashed white lines mark the grain boundaries. The tendency of the domain walls to orient perpendicular to the grain boundary is highlighted in (e). The arrows in (f) indicate that the domain walls do not continue over the grain boundary.

This observation leads us to the conclusion that the domain structures within the different grains are largely independent of each other. This unlocks the possibility to engineer the ferroelectric domains by imposing 3D geometrical confinement via the microstructure, representing an additional degree of freedom not available in $R$MnO$_3$ single crystals and thin films. To confirm that the domain and domain wall behavior observed in our PFM images applies in all spatial directions and not only to the surface region, we record complementary cross-sectional images using focused ion beam (FIB) in combination with scanning electron microscopy (SEM). The respective data is presented in Figure S4 (Supporting Information), showing the same features as observed at the surface, that is, vortex structures which unfold into stripe-like domains towards the sub-surface grain boundaries and domain walls with a preferred orientation perpendicular to them.

## 3. Inverse grain-size-dependent domain scaling behavior

In the next step, we systematically investigate the effect of 3D geometrical confinement on the domain structure by analyzing ErMnO$_3$ samples with different grain sizes manufactured using varying heat-treatment conditions as explained in the Supporting Information. To suppress unwanted cooling-rate-dependent variations in the domain structure as studied in refs. [45, 46, 62], the cooling rate is kept constant to 5 °C/min in all manufacturing processes. SEM micrographs (Figure S1, Supporting Information) indicate that we can readily tune the grain size in polycrystalline ErMnO$_3$ over a range of approximately 1.5 to 19.0 μm using heat-treatment conditions within the temperature range of 1350°C—1450°C ($T_C^{\text{ErMnO}_3} = 1150°C$).[62] An overview of the resulting ferroelectric domain structures for four samples with different grain sizes is presented in **Figure 2**. Analogous to the sample in Figure 1, a pronounced PFM contrast (out-of-plane) is observed for all samples showing the distribution of $\pm P$ domains. A closer inspection of the PFM data shows that polycrystalline ErMnO$_3$ samples with mean grain sizes from 4.8 to 19.0 μm develop qualitatively equivalent domain patterns, exhibiting a mixture of vortex and elongated stripe-like domains similar to those presented in Figure 1. For further reduced grain sizes, i.e., a mean grain size of 1.5 μm, a completely different behavior is observed. Here, the domain size approaches the grain size with no indication of either the vortex patterns or stripe-like domains found in the samples with larger grains.

To quantify the confinement-related variations in the domain structure, we analyze the domain size as function of the grain size. **Figure 3** displays the measured domain size in relation to the grain size of the polycrystalline ErMnO$_3$ samples synthesized under different heat-treatment conditions. To reliably quantify the domain size, we apply two different approaches, namely the maximum ball method and a stereographical method as explained in Figure S5 (Supporting Information).[63-65] Independent of the evaluation method, we find that the domain size decreases with increasing grain size; in other words, the largest grains exhibit the smallest domains. By fitting the universal scaling law (Equation (1)) to the data in Figure 3, we find a negative exponent $m = -0.1$. This scaling behavior is surprising as it is opposite to that known from classical polycrystalline ferroelectrics.[15-16] The latter is strikingly reflected by the literature data for $BaTiO_3$ ($m = +0.5$), which is plotted for comparison along with our $ErMnO_3$ results in Figure 3.[16, 18, 23, 66-68] The unusual domain scaling behavior in polycrystalline $ErMnO_3$ points towards a different and so far unknown origin, going beyond the domain-wall related reduction of elastic energies[25] that explains the positive scaling coefficients in, e.g., $BaTiO_3$[16] and $Pb(Zr,Ti)O_3$[15]. Note that in contrast to these materials, $ErMnO_3$ is co-elastic (not ferroelastic) and, hence, cannot simply release elastic strain by inserting or removing domain walls.[69]

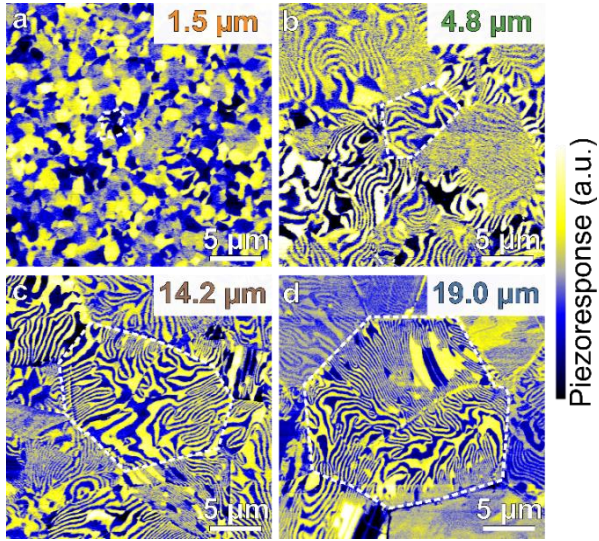

**Figure 2.** (a-d) Domain structures observed in samples with different grain size, obtained by different heat treatment conditions ((a) 1350°C, 10 min, (b) 1350°C, 4 hrs, (c) 1400°C, 12 hrs, (d) 1450°C, 12 hrs). PFM images (out-of-plane contrast) gained with voltage amplitude of 10 V at a frequency 40.13 kHz reveal qualitatively equivalent domain patterns for mean grain sizes of 4.8 μm, 14.2 μm, and 19.0 μm. In contrast, the domain size approaches the grain size for a mean value of 1.5 μm, leading to the suppression of vortex and stripe-like domains. For each mean grain size, the grain boundaries of one representative grain are highlighted by white dashed lines. SEM micrographs of the samples with different grain sizes are displayed in Figure S1 (Supporting Information).

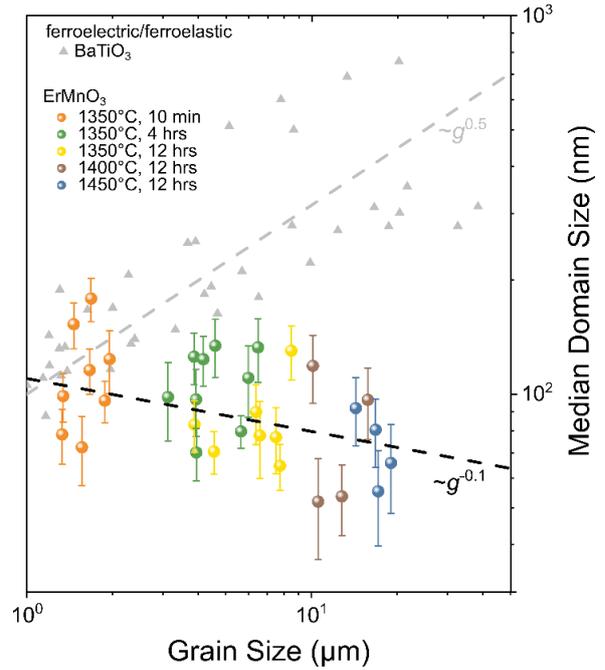

**Figure 3.** Grain-size dependence of the domain size in polycrystalline $ErMnO_3$. The domain size decreases with increasing grain size, corresponding to inverted domain scaling behavior with $d \sim g^{0.1}$ as indicated by the fit (black dotted line; a detailed description of the determination of the median domain size is provided in Figure S5, Supporting Information). For comparison, data for polycrystalline $BaTiO_3$[16, 18, 23, 66-68] is displayed, showing classical $\sim g^{0.5}$ scaling behavior (grey dashed line).

## 4. Strain-driven unfolding of vortex/anti-vortex pairs

In order to understand the origin of the inverse grain-size dependent domain scaling in Figure 3, we perform phase field simulations. The results are summarized in **Figure 4**. Analogous to previous studies,[27, 28, 30] the order parameter of the system is represented by a two-dimensional (2D) vector with its magnitude $Q$ and phase $\phi$ in our phase field model (see Supporting Information for details). Importantly, this approach allows for simulating the impact of an inhomogeneous strain field described by the distribution of the shear strain components, i.e., $\varepsilon_{xx} - \varepsilon_{yy}$, and $\varepsilon_{xy}$. The most pronounced strain-related feature arising from the geometrical 3D confinement in polycrystalline materials is the emergence of enhanced elastic strain at grain boundaries.[58, 59] In our simplified strain map, we account for this feature by introducing strain gradients as shown in Figure S6a (Supporting Information). Figure 4a and b show representative snapshots that document how the ferroelectric domain structure evolves under the inhomogeneous strain field. Colors indicate the six structural trimerization domains in $ErMnO_3$ labelled according to the color wheel in the inset to Figure 4a with $\alpha^+$, $\beta^+$, $\gamma^+$ corresponding to $+P$ domains and $\alpha^-$, $\beta^-$, $\gamma^-$ to $-P$ domains. A region with meandering domain walls and randomly distributed vortex/anti-vortex pairs arises in the central region of Figure 4b, corresponding to the established "zero-strain" ground state in hexagonal manganites.[40, 56-57] The vortex domains unfold into elongated stripe-like domains towards the boundaries with the domain walls oriented perpendicular to them. Furthermore, the simulations reveal that the relaxed domain structure forms via strain-driven annihilation of vortex/anti-vortex pairs, as well as vortex annihilation at the grain boundaries. The phase field simulations thus corroborate that the formation of the peculiar ferroelectric domain structures observed in polycrystalline $ErMnO_3$ is driven by elastic strain fields associated with the microstructural confinement.[58, 59]

To simulate the impact of varying grain size on the domain structure, we follow the assumption of strain fields with a constant decaying distance independent of the grain size.[70, 71] As displayed in Figures S6b (Supporting Information), in small grains, strain fields along the $x$ direction ($\varepsilon_{xx}$) and the $y$ direction ($\varepsilon_{yy}$) overlap, whereas the area where $\varepsilon_{xx}$ and $\varepsilon_{yy}$ do not overlap increases for larger grains. As shown by Equation S1 in the Supporting Information, the strain driving force is determined by the magnitude of $\varepsilon_{xx} - \varepsilon_{yy}$. Thus, in small grains where equally strong $\varepsilon_{xx}$ and $\varepsilon_{yy}$ overlap, the driving force for the unfolding of vortex/anti-vortex pairs vanishes. To test this hypothesis, we simulate the domain structures with grain-size independent strain fields. The domain structures are displayed together with the median domain size as a function of the box size in Figure 4c. As indicated by the dashed line, the median domain size decreases with increasing size of the simulated box, reproducing the trend observed in our experiments in Figure 3. However, we find that the negative scaling exponent extracted from the phase field simulations ($m = -0.4$) is larger than observed experimentally. Possible reasons for this discrepancy are the emergence of more complex grain-size dependent elastic strain fields[72, 73] and/or the presence of structural defects which can hinder the annihilation process of vortex/anti-vortex pairs. In summary, the simulations establish the correlation between emergent ferroelectric domain patterns and local elastic strain fields and provide a mesoscopic explanation for the anomalous grain-size dependent domain scaling behavior in polycrystalline ErMnO$_3$, revealing the annihilation of vortex/anti-vortex pairs as the key driving force.

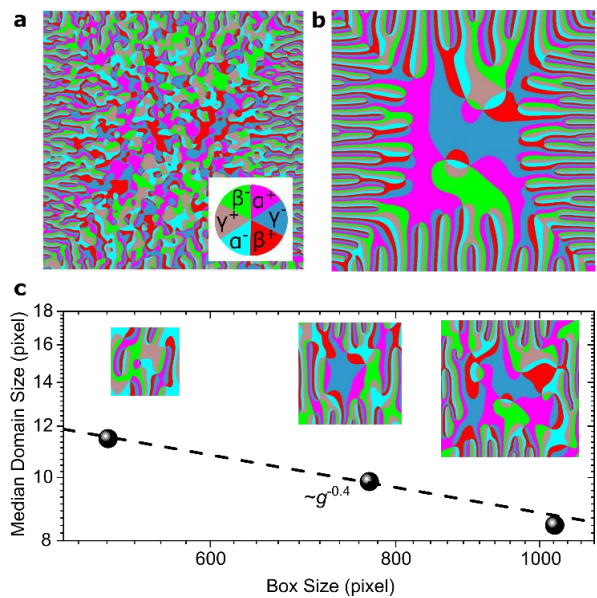

**Figure 4.** Interplay between elastic strain fields and vortex/anti-vortex pairs revealed by phase field simulations. Domain structures are simulated utilizing a gradient with enhanced elastic strain near the edge of the box (Figure S6a, Supporting Information).[58, 59] An (a) intermediate and (b) the final domain structure is displayed. The six structural trimerization domains are labeled according to the color wheel in the inset of (a) with $\alpha^+$, $\beta^+$, $\gamma^+$ corresponding to $+P$ domains and $\alpha^-$, $\beta^-$, $\gamma^-$ to $-P$ domains as displayed in the color wheel. (c) The median domain size is displayed as a function of the box size for superimposing strain fields (Figure S6b, Supporting Information). The domain size continuously decreases with increasing box size following the relationship $\sim g^{-0.4}$, which explains the inverted domain scaling behavior observed in our polycrystalline ErMnO$_3$ materials (Figure 3).

## 5. Conclusion and Outlook

Our results show that improper ferroelectric ErMnO$_3$ polycrystals exhibit inverted domain scaling behavior under 3D geometrical confinement compared to classical proper ferroelectrics, such as BaTiO$_3$[16] and Pb(Zr,Ti)O$_3$[15]. In contrast to the classical ferroelectrics, ErMnO$_3$ is not ferroelastic and, hence, cannot minimize its energy by forming additional strain-releasing domain walls. Instead, strain fields strongly interact with the topological vortices in ErMnO$_3$, promoting the annihilation of vortex/anti-vortex pairs and vortex annihilation at grain boundaries, which in turn determines the domain scaling. The substantial impact of topological defects on the grain-size dependent domain scaling behavior is intriguing as it provides a conceptually different handle for tuning the electromechanical and dielectric performance of ferroelectric materials, giving a new dimension to, e.g., macroscopic capacitor applications[74] and domain-wall based nanoelectronics[42-44]. The results are not restricted to ErMnO$_3$ and can be expected to arise in any material that naturally develops topologically protected defects, including hexagonal tungsten bronzes[75] and hexagonal ferrites[76]. Furthermore, recent advances in strain-field engineering utilizing dislocations,[11] secondary phases[12] and precipitates,[13] offer an interesting playground for future confinement studies. In general, designing polycrystals is a versatile processing-accessible path to adjust the domain structure of ferroelectrics utilizing topological defects, facilitating high tunability via strain fields, cooling rate,[45-47] and chemical doping[77] beyond the regimes accessible with single crystals and thin films.

### Acknowledgements

The authors acknowledge NTNU for support through the Enabling Nanotechnologies: Nano program. The Research Council of Norway is acknowledged for the support to the Norwegian Micro- and Nano-Fabrication Facility, NorFab, project number 245963/F50. J.S. acknowledges support from the Alexander von Humboldt Foundation through the Feodor-Lynen fellowship. J.S. and D.M. acknowledge NTNU Nano for the support through the NTNU Nano Impact fund. D.M. thanks NTNU for support through the Onsager Fellowship Program, the outstanding Academic Fellow Program, and acknowledges funding from the European Research Council (ERC) under the European Union's Horizon 2020 Research and Innovation Program (Grant Agreement No. 863691). H.W.Å acknowledges NTNU for financial support through the NTNU Aluminum Product Innovation Center. L.Q.C. acknowledges support by the Computational Materials Sciences Program funded by the U.S. Department of Energy, Office of Science, Basic Energy Sciences, under Award No. DE-SC0020145.

# Supporting Information

## 1. Synthesis and microstructural characterization of polycrystalline ErMnO$_3$

### 1.1. Synthesis of polycrystalline ErMnO$_3$

Synthesis of ErMnO$_3$ powder was done by solid-state reaction. The raw materials of Er$_2$O$_3$ (99.9% Purity; Alfa Aesar, Haverhill, MA, USA) and Mn$_2$O$_3$ (99.0% Purity; Sigma-Aldrich, St. Louis, MO, USA) were dried at 900°C and 700°C, respectively, for a dwell time of 12 hrs. Powders were mixed in a stoichiometric ratio and ball milled (BML 5, witeg Labortechnik GmbH, Wertheim, Germany) for 12 hours at 205 rpm using yttria stabilized zirconia milling balls with a diameter of 5 mm and ethanol as dispersion medium. After drying, annealing was done stepwise at 1000°C, 1050°C, and 1100°C for 12 hours. [1] Powders were mortared in between each firing step and the purity of the powders was checked using X-Ray Diffraction (XRD, D8 ADVANCE, Bruker, Billerica, MA, USA). Subsequently, the powder was pressed uniaxially into cylindrical pellets with a diameter of 10 mm, followed by isostatic pressing under a pressure of 200 MPa (Autoclave Engineers, Parker-Hannifin, Cleveland, OH, USA). Sintering was carried out in a closed alumina crucible with sacrificial powders of the same chemical compositions in the temperature range of 1350—1450°C with dwell times of 10 min—10 hrs (Entech Energiteknik AB, Ängelholm, Sweden) to produce samples with grain sizes in the range of 1.5—19.0 µm. Scanning electron microscopy (SEM) micrographs of polycrystalline ErMnO$_3$ manufactured under different sintering conditions are displayed in Figure S1. A heating and cooling rate of 5°C/min was utilized for all samples.

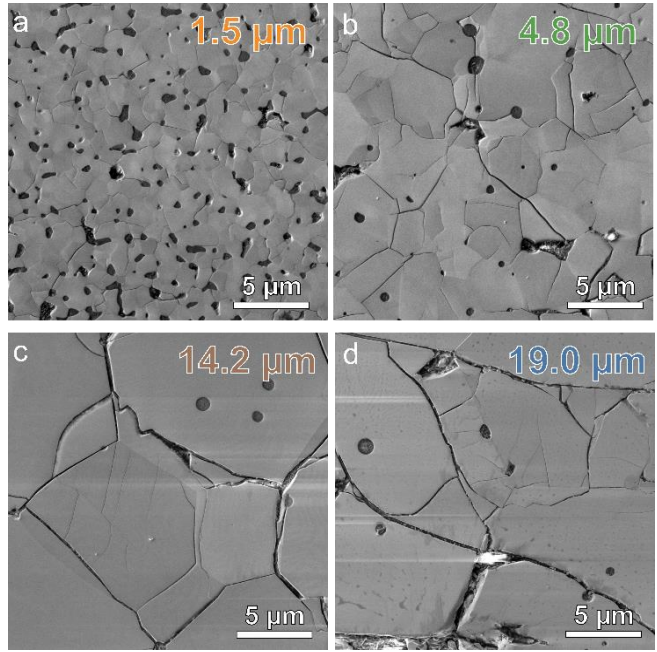

**Figure S1.** Impact of heat treatment on the microstructure evolution. SEM micrographs of polycrystalline ErMnO$_3$ sintered at (a) 1350°C, 10 min, (b) 1350°C, 4 hrs, (c) 1400°C, 12 hrs, and (d) 1450°C, 12 hrs.

As displayed in Figure 1 b-d and Figure S1, the microstructure of ErMnO$_3$ features a mixture of inter- and intragranular microcracking, commonly found in polycrystalline hexagonal manganites.[2] Microcracking originates from large thermal stresses induced during cooling, as discussed previously for polycrystalline YMnO$_3$ materials. Since microcracking occurs at much lower temperatures (620°C [3] in YMnO$_3$ polycrystals) than the formation of the domains at the Curie temperature ($T_c^{ErMnO_3}$~1150°C [4]), the observed domain structure is not affected by the microcracks.

### 1.2. Grain orientation maps

To obtain orientation maps (Figure 1b and Figure S2a), electron backscattered diffraction (EBSD) patterns were acquired on a NORDIF UF1100 detector in an Ultra 55 FEG-SEM (Zeiss, Jena, Germany) operated at 10 kV. The polished sample (heat treated at 1450°C, 12 hrs) was carbon coated prior to SEM imaging to enhance the conductivity. Patterns of 240 px x 240 px were recorded at a step size of 0.2 μm (six patterns per second) from a region of interest covering an area of (22.4 x 22.4) μm², featuring eight grains. Spatially averaged EBSD patterns of the different grains are displayed in Figure S2b. The patterns were indexed using dictionary indexing with EMsoft (version 5). [5] Orientations were uniformly sampled from the Rodrigues Fundamental Zone using Cubochoric sampling. [6] A total of 666187 patterns were simulated, utilizing a master pattern simulation of ErMnO$_3$ with the space group *P*6$_3$*cm*. [7] Orientations were refined by optimizing the match between experimental and simulated patterns utilizing EMsoft. Simulated patterns using the mean grain orientation after refinement are displayed in Figure S2c. Orientation analysis was done with MTEX 5.7 [8] and the respective orientations for the hexagonal unit cells are displayed in Figure S2d.

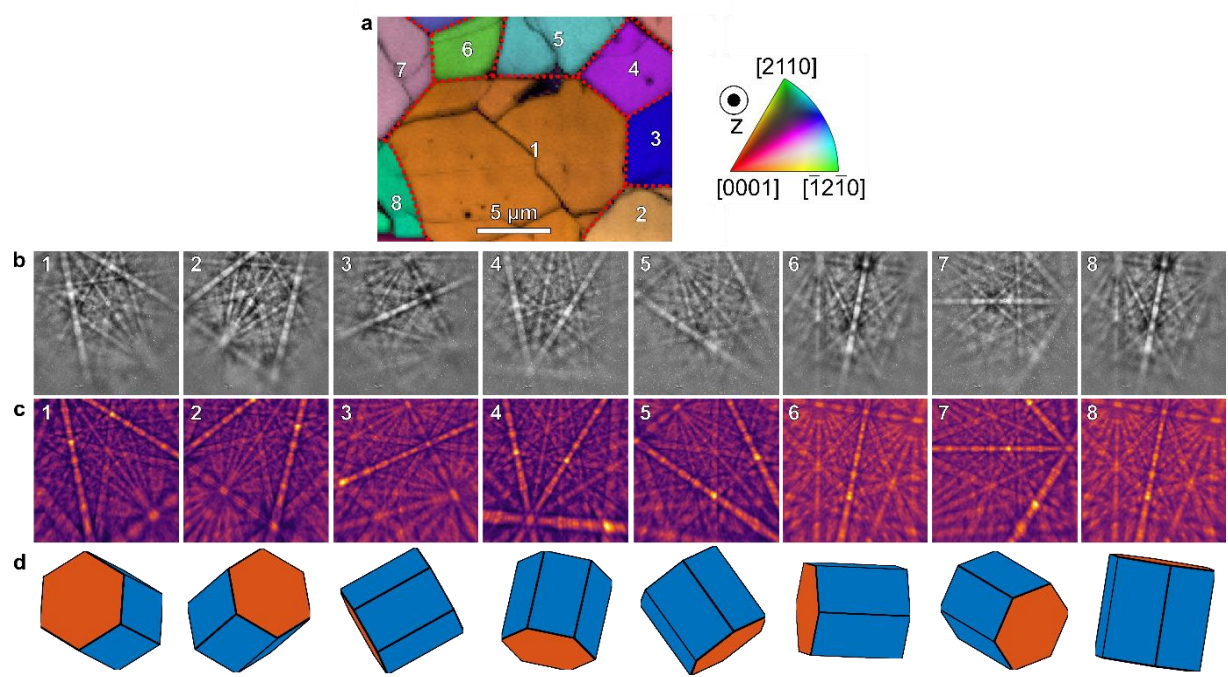

**Figure S2.** Analysis and fitting of EBSD data. (a) Orientation map as displayed in Figure 1b. (b) Spatially averaged EBSD patterns from the labeled 1-8 and (c) simulated patterns using the mean grain orientation. (d) The orientation of the individual grains is displayed schematically. Blue surfaces represent the $(10\bar{1}0)$ and red colors the $(0001)$ planes of the hexagonal crystal structure with space group *P*6$_3$*cm*.

## 1.3. Local electromechanical characterization

For local characterization, samples with a thickness of about 1 mm were lapped with a 9 µm-grained $Al_2O_3$ (Logitech Ltd, Glasgow, UK) water suspension and polished using silica slurry (Ultra-Sol® 2EX, Eminess Technologies, Scottsdale, AZ, USA) to produce a flat surface. Scanning probe microscopy measurements were performed on an NT-MDT Ntegra Prisma system (NT-MDT, Moscow, Russia). Scans were performed using an electrically conductive platinum tip (Spark 150 Pt, Nu Nano Ltd, Bristol, UK). For Piezoresponse Force Microscopy (PFM) measurements, the sample was excited using an alternating voltage (40.13 kHz, 10V). The laser deflection was read out by lock-in amplifiers (SR830, Stanford Research Systems, Sunnyvale, CA, USA). Prior to measurements on polycrystalline $ErMnO_3$, the PFM response was calibrated on a periodically out-of-plane poled $LiNbO_3$ sample (PFM03, NT-MDT, Moscow, Russia).

## 2. Domains and grain boundaries

### 2.1. Statistical evaluation of the domain wall/grain boundary angle

A statistical evaluation of the domain wall/grain boundary intersection angle is displayed for a representative grain (Figure 2d) in Figure S3. We find that more than 40% of the domain walls approach the grain boundary under an angle of 90°.

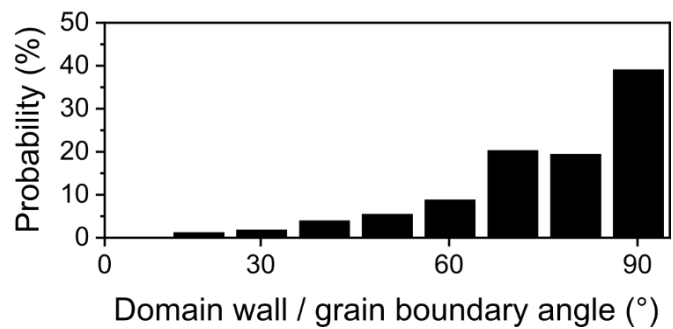

**Figure S3.** Statistical evolution of the domain wall/grain boundary angle.

## 2.2. Sub-surface domain structure

To investigate the 3D domain structure and its evolution towards sub-surface grain boundaries, we record cross sectional data (Figure S4a). To visualize the sub-surface domain structure, a polycrystalline $ErMnO_3$ sample (heat treated at 1350°C, 4 hrs) is loaded in a Helios Nanolab G4UX DualBeam focused ion beam (FIB) system (Thermo Fisher Scientific, Waltham, MA, USA), where SEM inspection of the sample surface confirms the presence of domain contrast,[9] matching the domain contrast imaged by PFM at the same position (Figure S4b). To enhance the conductivity of the surface of the sample, a thin layer of Pt/Pd (80wt.%/20wt.%, thickness: ~50 nm) is sputtered using a Cressington 208 HRB sputter coater (TESCAN GmbH, Dortmund, Germany). Next, following conventional preparation methods,[10] a cross section (depth ~15 µm) is milled out and polished with a 30 kV (90 pA) ion beam. At this point, the domain structure of the cross section of the lamella can be observed (Figure S4c). Combining the cross-sectional image with a PFM image at the same position allows for identifying the sub-surface domain wall structure in a grain of our polycrystalline $ErMnO_3$ sample. More details on the cutting procedure are provided in ref. [11].

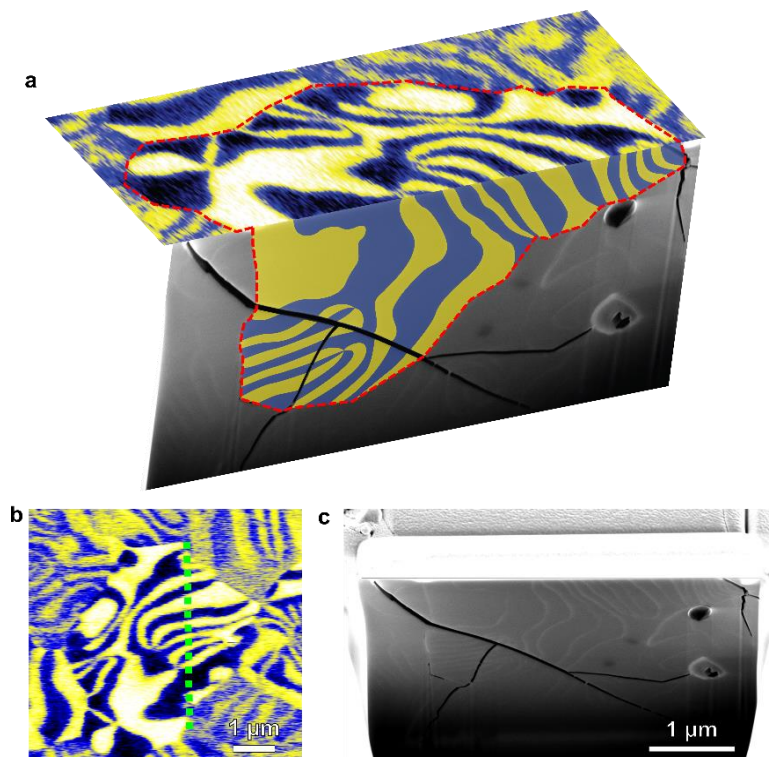

**Figure S4.** Cross sectional imaging of the subsurface domain structure of a grain of polycrystalline ErMnO$_3$. (a) Visualization of the surface and sub-surface domain structure of a grain of polycrystalline ErMnO$_3$ (sintered at 1350°C, 4 hrs). For a guide to the eye, we mark the domain structure in the cross section with an overlay. The dashed red line displays the position of the grain boundary. The surface domain structure is obtained from a PFM image, as displayed in (b). A FIB (as described in the method section and ref. [11]) is utilized to cut a cross section along the dotted green line. (c) A SEM image of the cross-section indicates the presence of domain wall contrast,[9] revealing the domain structure in the sub-surface region of the sample.

3. **Domain size analysis**

To analyze the grain-size dependence of the domain size, 33 grains of polycrystalline samples heat treated under different conditions were analyzed. To quantify the domain size, first, the PFM images were binarized. Domain contrast in an exemplary grain is displayed in Figure S5a, while the binarized image is displayed in Figure S5b. The individual domains with a difference of 180° in polarization orientation can be identified by black and white areas. In the post segmentation processing, a mean filter was utilized for dead-pixel correction (Figure S5c). Euclidean distance maps and skeletonized images of the binarized image are displayed in Figure S5d and e, respectively. Each pixel in the distance maps quantifies the distance of the pixel to the closest domain boundary. Figure S5f represents a mathematical combination of the Figures S5d and e (AND operation), yielding a skeletonized representation of the domain structure with pixel intensity corresponding to the distance to the closest domain boundary. Figure S5f shows this map with a (yellow) circle drawn centered on each pixel, with radius corresponding to the distance to the closest domain boundary. This corresponds to fitting a ball of maximum size[12] in the skeletonized image (Figure S5e). This representation replicates the domain structure of the initial PFM images well (Figure S5a). The median value of the radius of the fitted circles is used to quantify the domain size in polycrystalline ErMnO$_3$, which we display in Figure 3. To corroborate our results and to take the complex domain structure of our polycrystalline ErMnO$_3$ materials into account, a stereographical method[13] is used in addition. As explained in Figure S5g and h, the intersection of test lines with domains are counted and the domain width is quantified as $d = \frac{2}{\pi} \cdot \frac{\text{Total test line length}}{\text{Number of intersections}}$.[13] To exclude errors related to test lines oriented parallel to stipe-like domain walls, we rotate the image with respect to the test line in 10° steps (0°-90°) and average over the results.

The mean domain size obtained from the maximum ball method and the stereographical method are displayed in Figure S5i as a function of the grain size for all grains analyzed. The grain size was estimated from the area of the grain, assuming a circular grain morphology. The data is fitted (fits are displayed as dashed lines in Figure S5i) using the function $d = a \cdot g^m$, where $a$ and $m$ are fitting parameters. We find qualitatively similar $m$ values for the stereographical method ($m = -0.10$) and for the maximum ball method ($m = -0.14$).

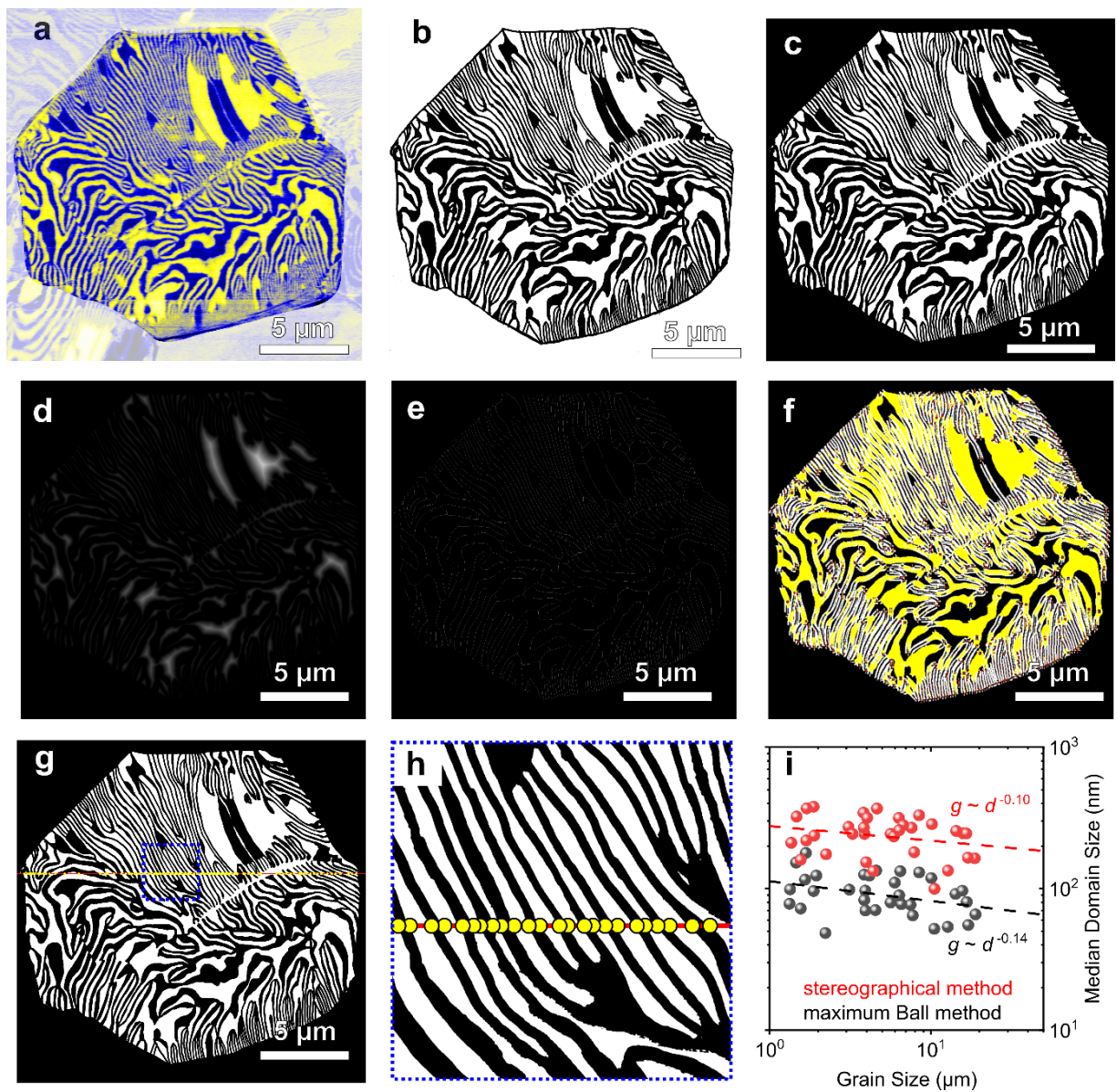

**Figure S5.** Methodology to determine the scaling behavior of ferroelectric domains in polycrystalline ErMnO$_3$. (a) PFM image displaying the grain of interest. (b) Binarized and (c) post-processed image corrected for dead pixels. The Euclidean distance map and the skeletonized image of (c) are displayed in (d) and (e), respectively. Mathematically combining the figures (d) and (e) results in the image displayed in (f). The image in (f) describes the domain structure of (a) utilizing circles (maximum ball method[12]). To corroborate our results, in addition, a stereographical method[13] is used to evaluate the domain size. As schematically explained in (g) (a magnification of the blue area is provided in (h)), the intersections (displayed as yellow dots) of the domains with test line (displayed in red) are counted. The median domain size, $d$, obtained from the stereographical (red) and the maximum ball method (black) is displayed as a function of the grain size, $g$, in (i). The dashed lines display fits to the experimental data.

## 4. Phase field modelling

In the phase-field simulations, the structural trimerization in ErMnO3 is described by the Cartesian coordinates $(Q_x, Q_y)$ and the polarization is described by $P_z$.[14, 15] Including the interaction between the order parameters and strain field, the free energy density is given by[14, 16]

$$
\begin{aligned}
f =& \frac{a}{2}(Q_x^2 + Q_y^2) + \frac{b}{4}(Q_x^2 + Q_y^2)^2 + \frac{c}{6}(Q_x^2 + Q_y^2)^3 + \frac{c'}{6}(Q_x^6 - 15Q_x^4Q_y^2 + 15Q_x^2Q_y^4 - Q_y^6) \\
& - g(Q_x^3 - 3Q_xQ_y^2)P_z + \frac{g'}{2}(Q_x^2 + Q_y^2)P_z^2 + \frac{a_P}{2}P_z^2 + \frac{s_Q^x}{2}[(\frac{\partial Q_x}{\partial x})^2 + (\frac{\partial Q_x}{\partial y})^2 + (\frac{\partial Q_y}{\partial x})^2 + (\frac{\partial Q_y}{\partial y})^2] \\
& + \frac{s_Q^z}{2}[(\frac{\partial Q_x}{\partial z})^2 + (\frac{\partial Q_y}{\partial z})^2] + \frac{s_P^z}{2}(\frac{\partial P_z}{\partial z})^2 + \frac{s_P^x}{2}[(\frac{\partial P_z}{\partial x})^2 + (\frac{\partial P_z}{\partial y})^2] - E_z P_z - \frac{1}{2}\varepsilon_0 \kappa_b E_z E_z \\
& + \lambda[(\varepsilon_{xx} - \varepsilon_{yy})(Q_x \frac{\partial Q_y}{\partial x} - Q_y \frac{\partial Q_x}{\partial x}) - 2\varepsilon_{xy}(Q_x \frac{\partial Q_y}{\partial y} - Q_y \frac{\partial Q_x}{\partial y})],
\end{aligned}
\quad (S1)
$$

where $a, b, c, c', g, g'$, and $a_P$ are coefficients for the Landau free energy function, $s_Q^x, s_Q^z, s_P^x$, and $s_P^z$ are coefficients for the gradient energy terms, $\varepsilon_0$ is the vacuum permittivity, $\kappa_b$ is the background dielectric constant[17], and $E_z$ is the electric field calculated by $E_z = -\frac{\partial \varphi}{\partial z}$ with $\varphi$ the electrostatic potential. Only $a$ is dependent on temperature, i.e., $a = a_0(T - T_C)$ with $T_C$ ~1150 °C[4]. The coefficients for the Landau free energy and gradient energy are obtained from first-principles calculations.[14] The temperature of the system is chosen as $T = 1149$ °C.

The system is evolved by solving the time-dependent Ginzburg-Landau (TDGL) equations

$$
\frac{\delta P_z}{\delta t} = -L_P \frac{\partial F}{\delta P_z}, \quad \frac{\delta Q_x}{\delta t} = -L_Q \frac{\partial F}{\delta Q_x}, \quad \frac{\delta Q_y}{\delta t} = -L_Q \frac{\partial F}{\delta Q_y} \quad (S2)
$$

where $F = \int f dV$, and $L_P$ and $L_Q$ are the kinetic coefficients related to the domain wall mobility. The TDGL equations are solved based on a semi-implicit spectral method, [18] and it is assumed that $L_P = L_Q = 0.05$ [a.u.].[19] The gradient energy coefficients are normalized based on $s_Q^{*z} = s_Q^{*z}/g_0$, $s_Q^{*x} = s_Q^{*x}/g_0$, $s_P^{*z} = s_P^{*z}/g_0$, and $s_P^{*x} = s_P^{*x}/g_0$, where $g_0 = a(\Delta x)^2$. The system size is $512\Delta x \times 1\Delta x \times 512\Delta x$, $768\Delta x \times 1\Delta x \times 768\Delta x$, and $1024\Delta x \times 1\Delta x \times 1024\Delta x$ with $\Delta x = 0.30 nm$. Periodic boundary conditions are applied to the system. To describe the effect of thermal fluctuations, small random noises are added to the order parameter components in the first 500 time steps, and all systems are evolved for a total of 90,000 time steps.

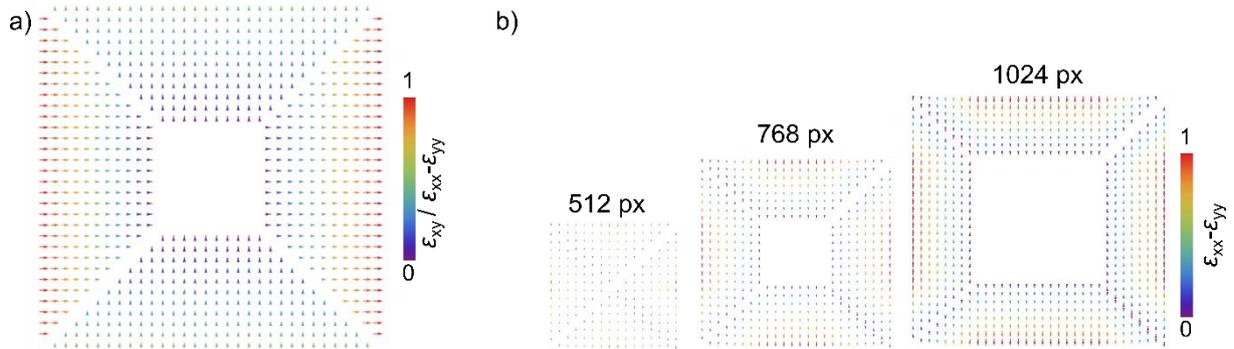

**Figure S6.** Simplified strain fields, representing the most prominent 3D confinement effects found in polycrystalline ceramic materials, which are enhanced strains at the grain boundaries and a strain free region in the center of the grain.[20] In a), the upper and lower boundaries possess $\varepsilon_{xx}$-$\varepsilon_{yy}$, while the left and right boundaries possess $\varepsilon_{xy}$. To simulate the impact of grain size, we assume for the strain fields in b), that the constraint from the neighboring grains is perpendicular to the grain boundaries, i.e., distribution of $\varepsilon_{yy}$ near the top/bottom boundaries and distribution of $\varepsilon_{xx}$ near the left/right boundaries.